\documentclass[aps, pra, reprint, superscriptaddress]{revtex4-1} 

\usepackage{amsmath, array} 
\usepackage{graphicx}
\usepackage{color}
\usepackage[normalem]{ulem}
\usepackage{soul,xcolor}
\usepackage{physics}
\usepackage{lineno}
\usepackage[toc,page]{appendix}

\usepackage{braket}

\newcommand{\EqLabel}[1]{\label{#1}} 

\newcommand{\John}{}

\begin{document}

\title{Bipolarons bound by repulsive phonon-mediated interactions}
\author{John Sous}  \affiliation{\!Department \!of \!Chemistry, 
  \!University of\!  British Columbia, \!Vancouver, British
  \!Columbia,\! Canada,\! V6T \!1Z1}  \affiliation{\!Department \!of \!Physics and
  Astronomy, \!University of\!  British Columbia, \!Vancouver, British
  \!Columbia,\! Canada,\! V6T \!1Z1}  \affiliation{\!Stewart Blusson Quantum Matter \!Institute, \!University
  of British Columbia, \!Vancouver, British \!Columbia, \!Canada,
  \!V6T \!1Z4}
\author{Mona Berciu} \affiliation{\!Department \!of \!Physics and
  Astronomy, \!University of\!  British Columbia, \!Vancouver, British
  \!Columbia,\! Canada,\! V6T \!1Z1}  \affiliation{\!Stewart Blusson Quantum Matter \!Institute, \!University
  of British Columbia, \!Vancouver, British \!Columbia, \!Canada,
  \!V6T \!1Z4}
  \author{Roman V. Krems} \affiliation{\!Department \!of \!Chemistry, 
  \!University of\!  British Columbia, \!Vancouver, British
  \!Columbia,\! Canada,\! V6T \!1Z1}
  

  
\begin{abstract}
When dressed particles (polarons) exchange quantum phonons, the resulting interactions are generally attractive. If the particles have hard-core statistics and the coupling to phonons is through the kinetic energy terms, phonon-mediated interactions are repulsive. Here, we show that such repulsive phonon-mediated interactions bind dressed particles into bipolarons with unique properties. These bipolaron states appear in the gap between phonon excitations, above the two-polaron continuum.  While thermodynamically unstable, the bipolaron is protected by energy and momentum conservation and represents a novel quasiparticle with a large dispersion and a negative effective mass near zero momentum. We discuss possible experimental implementation of the conditions for the formation of such repulsively bound bipolarons.
\end{abstract}
\date{\today}


\maketitle

\section{INTRODUCTION}
Most composite objects in nature are bound by conservative forces, which are attractive. Dissipative forces act to reduce the stability of bound objects by allowing tunnelling out of bound states, except in specific cases when dissipation traps the system in a local minimum of a conservative potential \cite{Lemeshko}.  For particles in a lattice, the dispersion is finite and the two- or few-particle continuum is bounded from both below and above. As a consequence, such particles can be bound by conservative repulsive interactions, which push the bound state to an energy above the continuum \cite{RepulsivelyBound}. The influence of such ``repulsively bound'' states on quantum walks of interacting bosons was recently demonstrated in experiments probing the dynamics of ultracold atoms in an optical lattice \cite{Greiner}. Repulsively bound states are fundamentally important as (i) they are expected to restrict certain quantum phases of many-body quantum systems to a finite range of Hamiltonian parameters \cite{[{For example, }]Phase}; (ii) they can induce correlations between particles in the high energy part of the continuum through virtual excitations; (iii) they are thermodynamically unstable states, which can nevertheless trap quantum systems thus impeding thermalization.  


Here we demonstrate the binding of particles by {\em repulsive} phonon-mediated interactions in the Peierls/ Su-Schrieffer-Heeger (SSH) model 
\cite{SSH_Barisic, SSH_Barisic_second, SSH_Barisic_third, SSH_PRL, SSH_RMP}. This problem is unique for two reasons. Firstly, phonon-mediated interactions in the celebrated Holstein and Fr{\"o}hlich models are generally attractive \cite{Bonca_bipolaron, Bonca_extended_bipolaron, Froh3, Froh4}. However, as we showed recently, the interactions between hard-core particles induced by coupling to quantum phonons described by the less studied Peierls/Su-Schrieffer-Heeger (SSH) model are repulsive \cite{Repulsive}. Secondly, the spectrum of phonons is unbounded from above. Any bound state embedded in the continuum of phonons should be expected to decay through coupling to phonons. However, if the phonons are gapped and the dispersion of the bound state is smaller than the gap, this decay may be prohibited by conservation of energy.  Under such conditions, the continuum of two particles $+$ phonon states separates into bands and the repulsive phonon-induced interactions lead to the formation of a stable bound state of two bare particles dressed by phonons, {\it i.e.} a bipolaron \cite{BipolaronsReview} pushed to an energy between the bands.  The repulsive bipolaron is particularly relevant to ultracold quantum simulators for which tunable gapped phonons can be engineered.

\begin{figure}[t]
  \centering \includegraphics[width=\columnwidth]{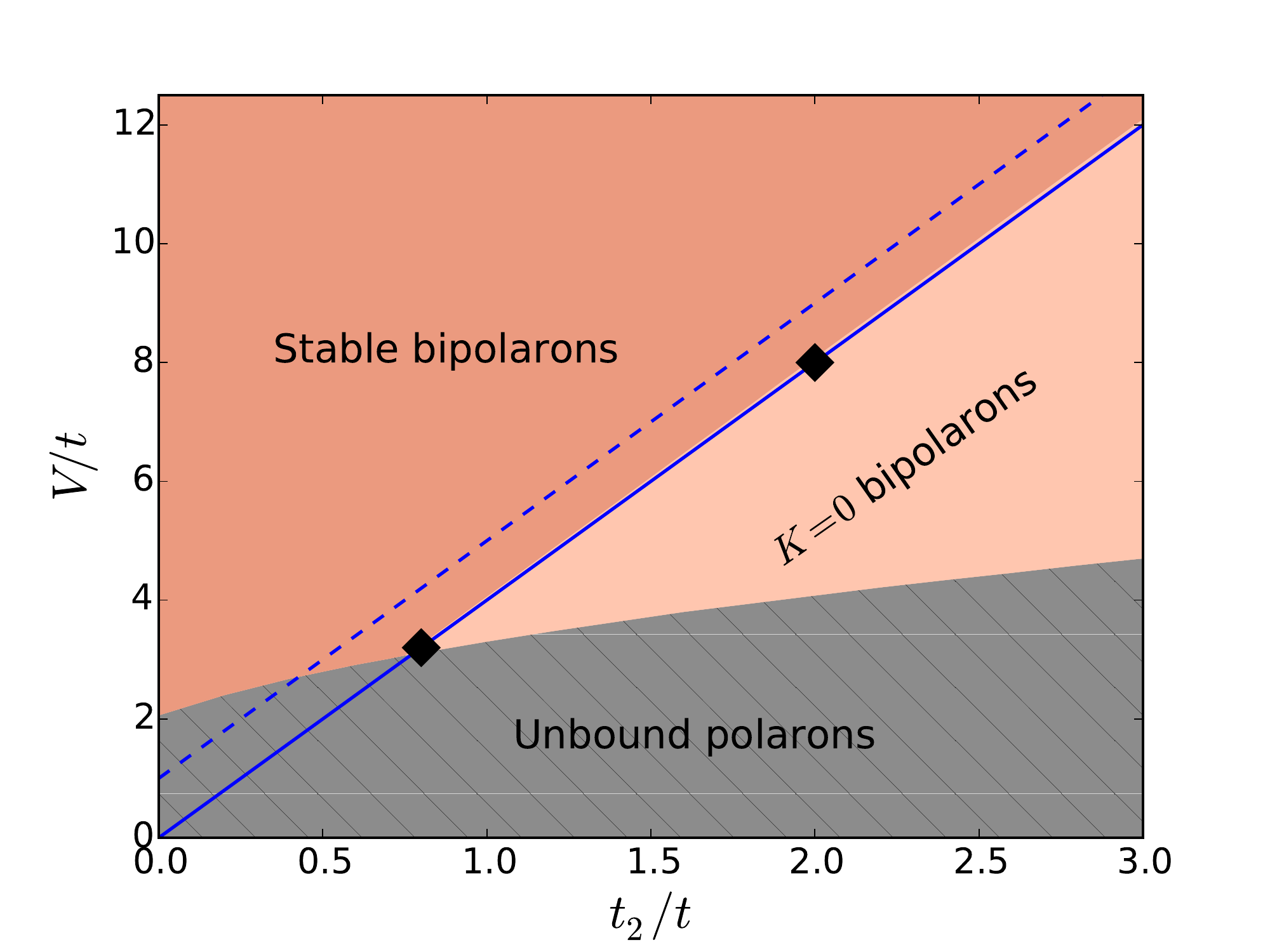} \caption{(color
    online) $V-t_2$ phase diagram. $t$ is the bare NN hopping.  The grey shaded region represents unbound polarons, while the light and dark salmon colored regions represent $K=0$ bipolarons and stable bipolarons, respectively. The blue solid line corresponds to the Peierls/SSH model for which $V = 4t_2$. The Peierls/SSH model line is right on the boundary between the region of stable bipolarons and the $K = 0$ bipolarons at $t_{2} = t_2^*\approx 0.8$ corresponding to $\lambda_C \approx 1.6$; the value of $\lambda$ marking the onset of stable Peierls/SSH bipolarons, where $\lambda = 2g^2/\hbar \Omega t$. The blue dashed line corresponds to the Peierls/SSH model supplemented with one unit of NN repulsion between the bare particles. The black diamond symbols mark the two points $\lambda = 1.6$ and $\lambda = 4$ on the Peierls/SSH line for which splitting of the repulsive bipolaron states from the continuum is illustrated in Fig. \ref{fig2}. \label{fig1}} \end{figure}

In general, dressing of particles by a bosonic or fermionic field leads to a variety of rich effects, ranging from polaronic effects \cite{Landau, Feynman1, Feynman2, Holstein1, Holstein2, Froh1, Froh2, Dominic} and BCS \cite{CooperPair,BCS1, BCS2} pairing in the dilute filling limit to Peierls and Mott physics at finite fillings \cite{hohenadler_review}. Polarons and bipolarons represent a classical problem in quantum field theory \cite{LLP, Migdal, Eliash}, and have acquired much recent attention due to experimental advances in atomic, molecular and optical (AMO) physics, with realizations of polaron continuum models as impurities in BECs \cite{bec3,bec8,bec9,bec12,bec13} and Fermi liquids \cite{fdg1,fdg2,fdg3,fdg4,fdg5} and polaron lattice models with cold atoms dressed by Rydberg excitations \cite{rydbergs3} and polar molecules \cite{polar-molecules1,polar-molecules2} in optical lattices, self-assembled ultracold dipolar crystals \cite{dipolar-crystals1,dipolar-crystals2a,dipolar-crystals2,dipolar-crystals3}, ions in rf traps \cite{ions2} and superconducting qubits \cite{d-wave1,d-wave3,d-wave4,d-wave7}. The interest in polaron physics stems from a fundamental research goal: understanding renormalization of quasiparticles and emergent interactions in quantum field theory. In most of these studies the focus is on ground state properties. Excited states are, however, a part of a rich and complex spectrum, where interactions between dressed particles are much less studied and understood. To this end, stable bipolarons in the spectral region above the two-polaron continuum represent a new paradigm for binding of dressed particles.


\section{MODEL}
We consider the Peierls/Su-Schrieffer-Heeger(SSH) model for hard-core particles ({\em e.g.}, spinless fermions, hard-core bosons) on a one-dimensional lattice, ${\cal H} = {\cal H}_{\rm p} + {\cal H}_{\rm ph} + V$, where
\begin{eqnarray}
{\cal H}_{\rm p} =-t \sum_{i}^{}\left( c_i^\dagger c_{i+1} +
h.c.\right)
\end{eqnarray}
is the tight-binding model of the bare particles with nearest-neighbour (NN) hopping ($h.c.$ is the Hermitian conjugate) for which $\left \{c_i,c_j^\dagger \right \}_\pm = \delta_{ij}$, $\left \{c_i,c_j \right \}_\pm = \left \{c_i^\dagger,c_j^\dagger \right \}_\pm= 0$ and $i$ is the site index. The subscript $\pm$ refers to the anticommutator for fermions and the commutator for bosons, respectively.  We enforce the hard-core condition $c_i^\dagger c_i^\dagger = 0$.  ${\cal H}_{\rm ph}$ is given by
\begin{eqnarray}
{\cal H}_{\rm ph} = \hbar \Omega \sum_{i}^{} b_{i}^\dagger b_i
\end{eqnarray}
with $[b_i,b_j^\dagger]=\delta_{ij}$, $[b_i,b_j]=[b_i^\dagger,b_j^\dagger]=0$.  This term describes Einstein phonons of frequency $\Omega$, {\em i.e.} an infinite ladder of states separated by $\hbar \Omega$. The interaction
\begin{equation}
\EqLabel{phonon-coupling} \hat{V}=g\sum_{i}^{}\left( c_i^\dagger c_{i+1} +
h.c.\right)\left( b_i^\dagger+b_i - b_{i+1}^\dagger-b_{i+1}\right)
\end{equation}
is the Peierls/SSH particle-phonon coupling. This interaction describes the modulation of the hopping amplitude by out-of-phase ``breathing mode" phonons \cite{SSH_Barisic, SSH_Barisic_second, SSH_Barisic_third, SSH_PRL, SSH_RMP}. We characterize the particle-phonon coupling by the dimensionless parameter
\begin{equation} 
\lambda = 2g^2/\hbar \Omega t . 
\end{equation}

In this work, we focus on the anti-adiabatic limit $t, g \ll \hbar \Omega$ {\John{(as follows from the discussion below, the regime of particular interest here is $t \ll g\ll \hbar \Omega$)}} and investigate the conditions for the formation of repulsively bound bipolarons.  The repulsive bipolaron is separated from the higher bipolaron$+$phonon states by energy gaps that are proportional to $\Omega$.  Thus, the repulsive bipolaron is expected to be stable in this limit.


\section{EFFECTIVE HAMILTONIAN}

{\John{We set the lattice constant $a$ to unity in all the following equations.}} 

In the two-bare-particle sector, the SSH interaction (\ref{phonon-coupling}) induces both repulsive phonon-mediated density-density interaction and ``pair-hopping" interactions for particles with hard-core statistics \cite{Repulsive} as we explain below. 

In the anti-adiabatic limit, the contribution of $N$-phonon excitations decays as $(g/\hbar \Omega)^N$.  This allows us to derive an effective theory to order $(g/\hbar \Omega)$ by projecting out higher energy multi-phonon states. For more details, see Appendix \ref{PT}.
We note that this approach has been shown to be quantitively accurate in the anti-adiabatic limit \cite{Dominic, Repulsive} and furthermore elucidates the intricate physical mechanisms induced by interactions with phonons.  We will comment on the generality of our results in a following section.

We first consider a single particle coupled to phonons. This effective Hamiltonian first derived in Ref. \cite{Dominic} reads: 
\begin{eqnarray}
\hat{h}_1 = -\epsilon_0\sum_{i}^{}\hat{n}_i + \sum_{i}^{}( -t c_i^\dagger c_{i+1}+ t_2 c_i^\dagger c_{i+2} + h.c.),
\end{eqnarray}
where now the $c$ operators act in the space of polarons. This Hamiltonian describes a dressed particle characterized by the NN hopping $t$ and 
an effective phonon-mediated next-nearest-neighbour (NNN) hopping $t_2=g^2/\hbar \Omega=\lambda t/2$, arising from the interaction between the bare particle and the phonons.  The NNN hopping can be viewed as a second-order process: $c^\dagger_i|0\rangle
\stackrel{\hat{V}}{\Longrightarrow}c^\dagger_{i+1} b^\dagger_{i+1}|0\rangle
\stackrel{\hat{V}}{\Longrightarrow} c^\dagger_{i+2} |0\rangle$. 
Note that the NNN hopping must be positive $t_2 > 0$.

\begin{figure*}[t]
    \centering \includegraphics[width=1.0\columnwidth]{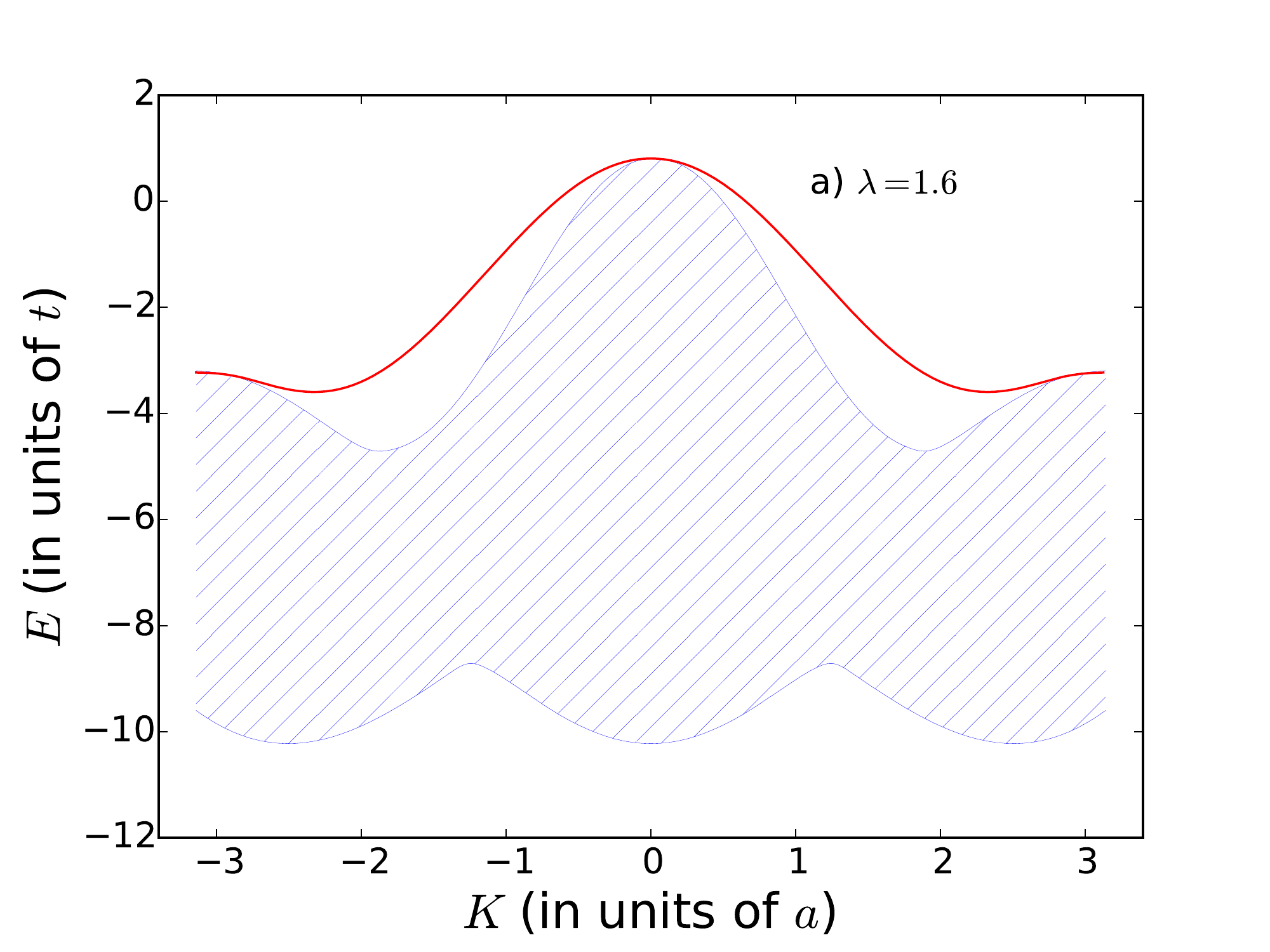} \centering
\includegraphics[width=1.0\columnwidth]{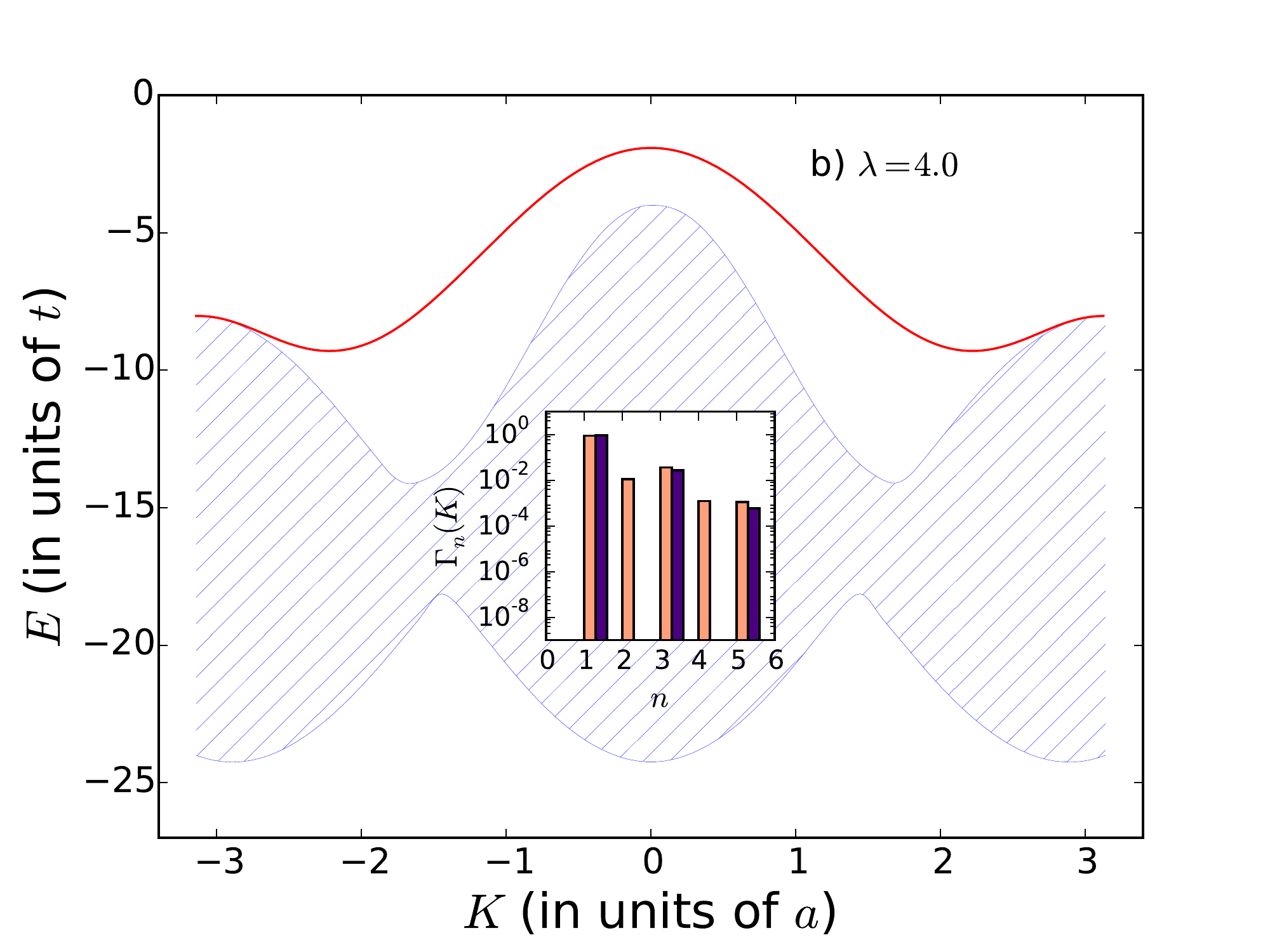} \caption{(color online)
Energy spectrum of two hard-core bare particles in the one-dimensional Peierls/SSH model.  The red line represents the repulsive bipolaron dispersion and the blue shaded region shows the two-polaron continuum band for a) $\lambda=1.6$ and b) $\lambda=4$, where $\lambda = 2g^2/\hbar \Omega t$, $t$ is the bare NN hopping and $a$ is the lattice constant. For $\lambda=1.6$, parts of the band are split from the continuum, while for $\lambda=4$ the whole band is split except near the edges of the Brillouin zone. The inset of b) illustrates the repulsive bipolaron log probability distribution $\Gamma_n(K) = log_{10}[P_{n}(K)]$ defined for $P_{n}(K) \equiv \abs{\braket{K,{BP}|K,n}}^2$, where $\ket{K,{BP}}$ is the bipolaron state and $n$ is the relative separation of the particles for $K_{BP} = 0$ (salmon) and $K_{BP} = \pi$ (indigo). In both cases, the particles are NN with highest probability. Note that for $K_{BP} = \pi$, even $n$ relative separation between particles is forbidden. For more details, see Appendix \ref{EOMBBGKY}. \label{fig2}} \end{figure*} 
 
This term leads to a reduced effective mass at strong coupling and thus indicates a departure from typical polaronic behavior of the Holstein and Fr{\"o}hlich models, where the coupling enhances the polaron's effective mass.  For more details, see Ref. \cite{Mona_SSHplusH}.
 
The four processes $c^\dagger_i|0\rangle
\stackrel{\hat{V}}{\Longrightarrow}c^\dagger_{i\pm 1} b^\dagger_{i\pm
1}|0\rangle \stackrel{\hat{V}}{\Longrightarrow} c^\dagger_{i} |0\rangle$ and
$c^\dagger_i |0\rangle\stackrel{\hat{V}}{\Longrightarrow}c^\dagger_{i\pm 1}
b^\dagger_{i}|0\rangle \stackrel{\hat{V}}{\Longrightarrow} c^\dagger_{i}
|0\rangle$ give rise to the polaron formation energy $\epsilon_0= 4g^2/\hbar \Omega=2\lambda t$.  The resulting polaron dispersion is 
\begin{eqnarray}
E_P(k)= - \epsilon_0 - 2t \cos (k) + 2 t_2 \cos (2k).
\end{eqnarray}

Here we are particularly interested in the case of $t_2 > t$, where the dispersion is dominated by the phonon-mediated interactions. 
Note that $t_2$ can exceed $t$ even if $g/\hbar \Omega \ll 1$. For a particular example, consider the case of $g/\hbar \Omega = 0.1$ and $g > 10{\John{~t}}$, yielding the desired result.   

Repeating the calculation for two particles \cite{Repulsive}, we find
\begin{eqnarray}
\hat{h}_2 =\hat{h}_1 + \epsilon_0 \sum_{i}^{}
\hat{n}_i\hat{n}_{i+1}, 
\label{repulsion}
\end{eqnarray}
illustrating the appearance of  phonon-mediated NN {\em repulsion}. Its origin can be explained as follows: if the particles are $n \ge 2$  sites apart, each lowers its energy by $\epsilon_0$ through hops to its adjacent sites and back, accompanied by virtual phonon emission and absorption, as explained above. However, if the particles are on adjacent sites, then the hard-core condition blocks half of these processes, {\it i.e.} each particle can only lower its energy by $\epsilon_0/2$. Thus, there is an energy cost for particles to be adjacent equal to $\epsilon_0=2\lambda t$.

This Hamiltonian also includes ``pair-hopping" interactions which mediate the hopping of NN pairs via exchange of phonons. In this process the pair moves as a whole as opposed to hopping of one particle past the other which is forbidden by the hard-core statistics. In the anti-adiabatic limit considered in this work, the NNN hopping precisely compensates for the pair-hopping process. See Ref. \cite{Repulsive} for more details.

Two Peierls/SSH polarons in the limit $g/\hbar \Omega \ll 1$ are thus described by the limiting case of 
the $t-t_2-V$ model
\begin{eqnarray}
{\cal H}_{\rm eff} &=& -\epsilon_0
\sum_{i}^{}\hat{n}_i - t \sum_{i}^{} \left(c_i^\dagger c_{i+1} + h.c.\right)  \nonumber \\
&+&  t_2 \sum_{i}^{}\left(c_i^\dagger c_{i+2} + h.c.\right) + V \sum_{i}^{} \hat{n}_i\hat{n}_{i+1}, 
\label{TheModel}
\end{eqnarray}
with {\John{$V = \epsilon_0 = 2 \lambda t$}} and $t_2 = \lambda t /2$.

This model is characterized by NNN hopping and pair-hopping for NN bound pairs both with a sign opposite to that of the NN hopping. Thus, this model cannot be taken as a limit to a long-range power law hopping model.  This is a unique feature of this model which explains why the repulsive bipolarons formed in this model are different from the $t_2 = 0$ repulsively bound pairs, as we shall demonstrate below.

To understand phonon-mediated pairing, we first study the full phase diagram of the model (\ref{TheModel}) in the entire range of $V$ and $t_2$. We derive an exact Bogoliubov-Born-Green-Kirkwood-Yvon (BBGKY) equation-of-motion (EOM) to solve for the two-particle propagator \cite{CiniSawatzky, MBmp} $G(K,n,\omega)=\bra{K,1}\hat{G}(\omega)\ket{K,n}$ defined for two-particle states $\ket{K,n} = \sum_i \frac{e^{iK(R_i+n/2)}}{\sqrt{N}}c_i^\dagger c_{i+n}^\dagger \ket{0}$ with $n \geq1$. We extract bound state properties from the pole of the propagator that appears above the continuum band. 
The continuum is the convolution of two single polaron bands satisfying the conservation of momentum.  We briefly outline the calculation procedure in Appendix \ref{EOMBBGKY}.

\section{RESULTS}
We set $\hbar =1$ and study the phase diagram of the model (\ref{TheModel}). Fig. \ref{fig1} shows the regions in parameter space characterizing the appearance of stable repulsive bipolarons, metastable repulsive bipolarons, and unbound polarons. We define stable bipolarons as those with energy separated from the continuum at all values of the bipolaron momentum $K$. By contrast, metastable bipolarons are defined as bipolarons with energy dispersion split from the continuum at $K = 0$, while merging with the continuum at other values of $K$. Such bipolarons may dissociate by momentum-changing collisions. 

\begin{figure}[t]
    \centering \includegraphics[width=1.0\columnwidth]{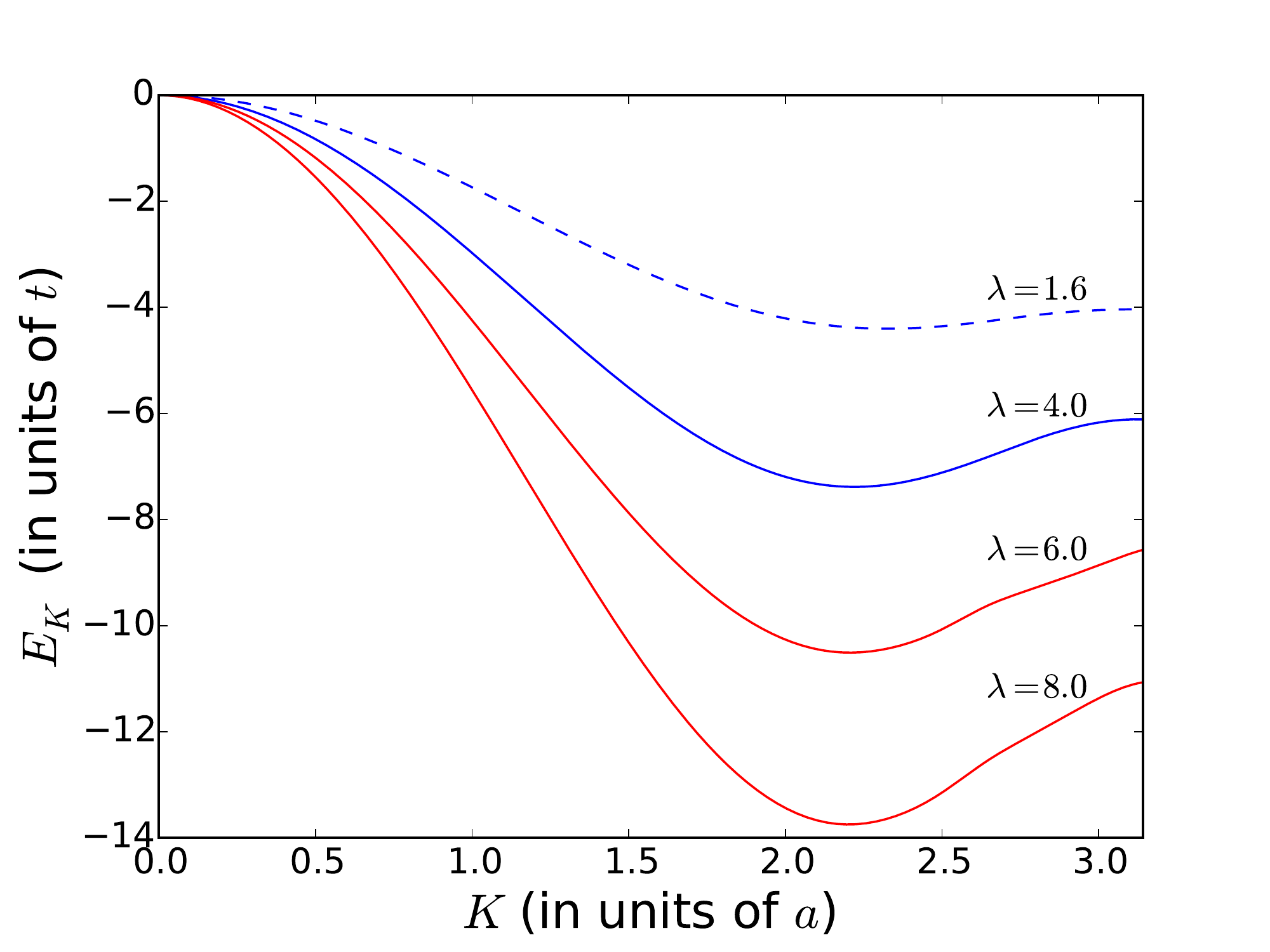} 
\caption{(color online)
The repulsive bipolaron dispersion $E_K \equiv E_{BP}(K) - E_{BP}(0)$ in units of the bare NN hopping, $t$, for various values of $\lambda = 2g^2/\hbar \Omega t$. $a$ is the lattice constant. The dashed line represents the onset of Peierls/SSH bipolaron formation. The two blue lines were illustrated in Fig. \ref{fig2}, while the red lines label strong coupling bipolarons. Note that the bipolaron dispersion exhibits both large curvature and bandwidth, which increase with the coupling strength. \label{fig3}}
 \end{figure}

To understand the binding mechanism, we first consider the effect of the $t_2$ term on the continuum band. 
As $\lambda$ increases, the NNN hopping dominates leading to a transition of the single polaron dispersion from one with a minimum at 
 $k = 0$ to one with a doubly degenerate dispersion with two minima at finite $k = \pm\pi/2$. This leads to an asymmetry in the two-polaron continuum with upward shifting near the center of the Brillouin zone above the zero of energy. Note that the asymmetry is reversed for $t_2 \rightarrow -t_2$ as the single polaron band would then have two maxima instead of two minima. Thus, this NNN hopping cannot be considered as a cut-off to a longer-range hopping as we argued before. As $t_2$ increases, the continuum band broadens near the bottom and narrows at the top and near the edges (not shown) till $|t_2|>|t|$, after which this asymmetry decreases (see Fig. \ref{fig2}) as the NNN hopping dominates and the NN hopping becomes a perturbative term.

\begin{figure}[t]
    \centering \includegraphics[width=1.0\columnwidth]{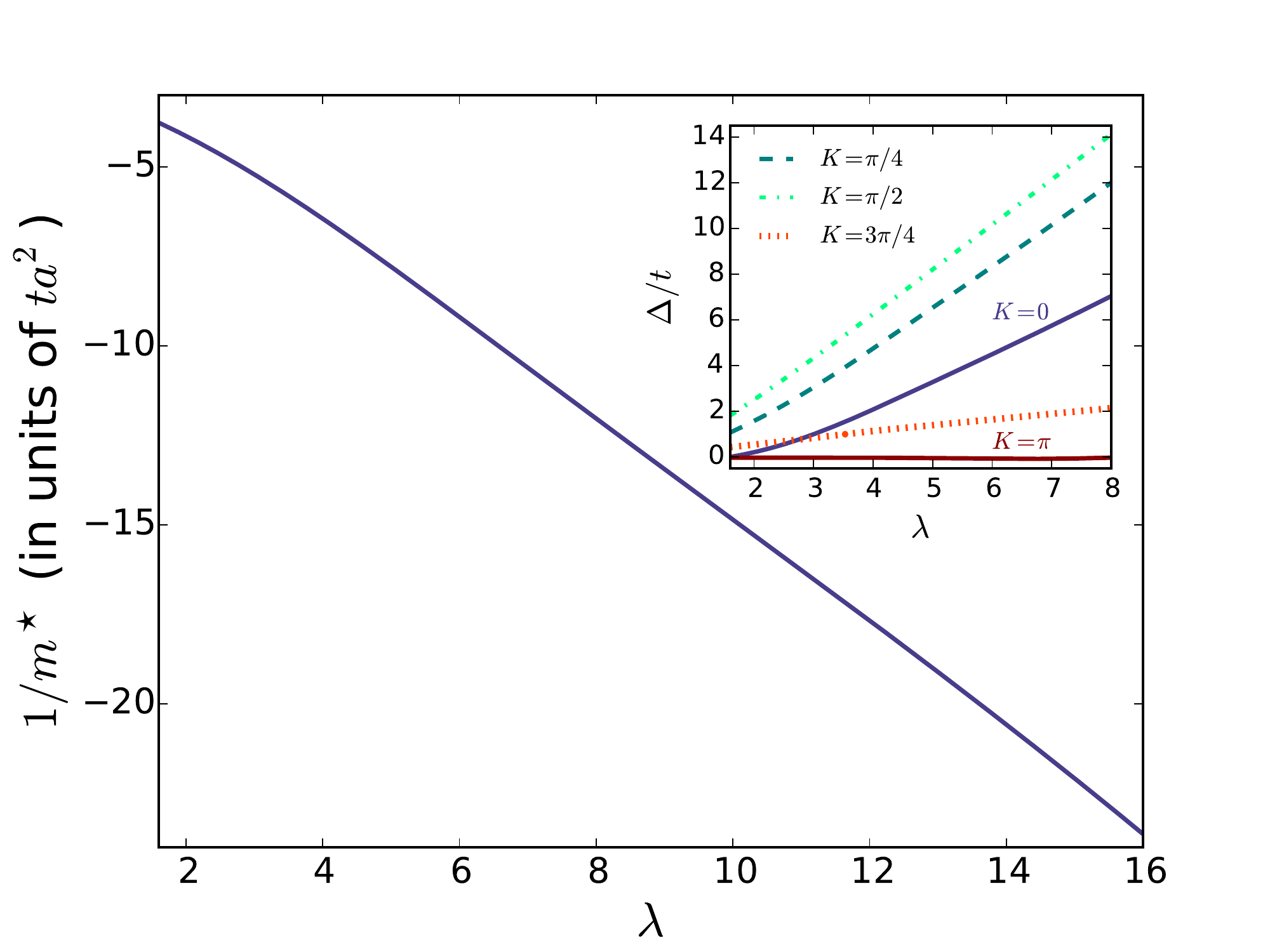} 
\caption{(color online)
The dependence of the inverse effective mass, ${ta^2}/{m^*}$, for the repulsive bipolaron on $\lambda = 2g^2/\hbar \Omega t$. The effective mass of the bipolaron $m^*$ is defined as $m^* = (\partial^2{E_{BP}(K)} / \partial{K}^2)^{-1}$ for ${K=0}$, $t$ is the bare NN hopping and $a$ is the lattice constant.  The inset illustrates the dependence of the energy gap, $\Delta$, between the repulsive bipolaron and the edge of the two-polaron continuum band on $\lambda$ at various values of $K$. The dark blue and dark red solid lines label the gaps for $K = 0$ and $K = \pi$, respectively; the dashed, dash-dotted and dotted lines label the gaps for $K = \pi/4$, $K = \pi/2$ and $K = 3\pi/4$, respectively. Note that $\Delta$ vanishes for $K = \pi$.   
 \label{fig4}}
 \end{figure}

To bind polarons, the interaction must compensate for the kinetic energy lost by binding. For $t_2=0$, binding happens for $V>2~t$ \cite{Vektaris}. However, $t_2$ enhances the kinetic energy of individual polarons shifting the continuum band center upwards as explained above. Thus, a higher $V$ is required to bind polarons. This is what we observe in numerical calculations shown in Fig. \ref{fig1}. 
 The $K=0$ bound states first appear at a critical value $t_2^*\approx 0.8~t$. For $t_2<t_2^*$, a bound state is split everywhere in the Brillouin zone, once it is split at $K = 0$. As $t_2$ surpasses $t_2^*$, the asymmetry is lost and the continuum edges shift upwards and become closer in energy to the continuum center. Thus, a greater $V$ is required to split the entire spectrum at higher $K$. This implies that metastable $K=0$ bipolarons can decay into the continuum near the Brillouin zone edges by collisions with other bipolarons or through other $K$-changing mechanisms. 

For the Peierls/SSH model, the ratio $V/t_2 = 4$ is fixed, as shown by the blue line in Fig. \ref{fig1}. When $t_2 > t_2^*$, the Peierls/SSH model line is right on the boundary between the region of stable bipolarons and the $K = 0$ bipolarons. 
Fig. \ref{fig2} demonstrates the splitting of repulsively bound states from the continuum for two points on this line marked by symbols in Fig. \ref{fig1}. 
Fig. \ref{fig2}a) corresponds to $\lambda = 1.6$ ($t_2 = t_2^*$), illustrating the onset a bound state at $K=0$. Fig. \ref{fig2}b) corresponds to $\lambda = 4$, illustrating the onset of a bound state split from the continuum at all values of $K$. This state merges with the continuum near the edge of the Brillouin zone. The $V/t_2 = 4$ line at $t_2 > t_2^*$ thus marks the onset of stable bipolarons. Note that any repulsion between bare particles contributes directly, pushing the energy of the bound state from the continuum. This is illustrated in Fig. \ref{fig1} by the dashed line, which corresponds to the Peierls/SSH model supplemented with one unit of NN repulsion between the bare particles.


The bipolarons illustrated here have interesting properties.  As can be seen from the dispersion of the bound state in Fig. \ref{fig2}, the bipolaron has negative effective mass at the band center and positive effective mass at the band minimum. Therefore, coupling the $K=0$ bipolarons to photons may result in negative refraction \cite{NegEmass1,NegEmass2,NegEmass3,NegEmass4}. 
Even more interestingly, the dispersion of the bipolaron exhibits large curvature, which {\it increases} with the coupling strength, as illustrated in Fig. \ref{fig3}.  

The unique shape of the bipolaron band in Fig. \ref{fig3} can be explained by considering the limit $V >> t, t_2$. The dispersion is then $ \sim [(2 t^2 / V) + 2t_2] \cos(K) +  (2t_2^2/V) \cos(2K)$.  The first term represents NN hopping of a bound pair comprised of two NN hops of the constituent polarons in the same direction with amplitudes $t^2/V$ and an additional pair-hopping $t_2$.  The second term represents NNN of a bound pair where one particle moves two sites away from the its bound partner by NNN hopping and then the partner follows, costing an energy $V$ in the process.  In the Peierls limit, $V= 4 t_2$, the analytical expansion acquires higher order corrections.  However, this simple analytical form provides a way to explain the unusual form of the bipolaron band.  In particular, the second term $\sim \cos(2K)$ is responsible for the unusual curvature of the dispersion with a minimum at finite $K < \pi$.  This analysis makes clear that the pair-hopping term is responsible for the bipolaron's enhanced band width and reduced effective mass.

To illustrate this quantitatively, we plot in Fig. \ref{fig4} the inverse of the bipolaron's effective mass $m^*$ as a function of $\lambda$.  We find that $|m^*|$ decreases rapidly with $\lambda${\John{, and consequently, with $V$, as $V = 2 \lambda t$ for the Peierls/SSH polarons.}}  For contrast, a repulsively bound pair in the bare $t-V$ model with $t_2 = 0$ has an effective mass $m_0 = -V/(2t^2)$ \cite{Vektaris}, which leads to linear growth of $|m_0|$ with $V$.  Furthermore, the energy gap between the $K = 0$ bipolaron and the two-polaron continuum, $\Delta$, grows with $\lambda$, as seen in the inset of Fig. \ref{fig4}.  This suggests that the $K= 0$ bipolaron is sufficiently stable at strong coupling.  This radically distinct behavior highlights the unique nature of the bipolaron which is expected to be highly mobile near $K = 0$. 

\section{Beyond the anti-adiabatic limit}
{\John{
In the previous section, we considered the anti-adiabatic limit.  At lower values of phonon frequency, the bipolaron acquires higher order vertex corrections.  This leads to quantitative changes in the results.  However, previous studies suggest that the qualitative physics in the anti-adiabatic limit persists at lower phonon frequencies \cite{Dominic, Repulsive}. 
In particular, Ref. \cite{Dominic} shows that the sharp transition of the ground state of a single SSH polaron from $k=0$ to finite $k$ occurs both in the adiabatic and anti-adiabatic regimes. The critical value of $\lambda$ corresponding to the transition varies within a small range between $0.5$ and $1.25$ as the phonon frequency is decreased from the anti-adiabatic limit to within the adiabatic regime. The polaron dispersion varies smoothly as the phonon frequency is decreased beyond the anti-adiabatic limit and exhibits no irregularities in both the anti-adiabatic and adiabatic limits. Since here we consider the same model, we expect a similarly smooth variation of the bipolaron dispersions beyond the anti-adiabatic limit. In fact, the novel features of the bipolarons considered here can be traced back to the interplay of the bare particle statistics and phonon-mediated hopping for the single SSH polaron.

For example, as discussed above, the repulsive interactions stem from the statistics blocking the phonon-mediated hopping of bare particles into the same lattice site, thereby eliminating part of the renormalization energy (polaron formation energy) of the individual polarons. Since the renormalization energy of the single polaron is a smooth function of phonon frequency, the repulsive interactions are expected to extrapolate smoothly beyond the anti-adiabatic limit. The result displayed in Figure \ref{fig4}, showing the decrease of the bipolaron mass with the coupling strength, is a consequence of pair-hopping, which is closely connected to the NNN polaron hopping.  The NNN polaron hopping is responsible for the decrease in the single polaron mass in the anti-adiabatic regime at strong coupling.  As Ref. \cite{Dominic} shows, the single SSH polaron becomes heavier as the phonon frequency is decreased but remains light well beyond the anti-adiabatic limit. The same should be expected for the repulsive bipolaron.

Note that we cannot argue that the results of the present work apply to the adiabatic regime. However, we use the above observations to suggest the validity of our qualitative results at phonon frequencies beyond the anti-adiabatic approximation, as long as $\hbar \Omega$ is larger than the bipolaron's bandwidth, the condition that ensures a gap between the bipolaron state and the higher continuum states.

To tackle this problem more rigorously, one may attempt to use numerically exact approaches.  However, typical large-scale computational techniques are limited to the ground state, while the repulsive bipolaron state of interest is an excited state.  One possible approach to circumvent this difficulty and access a solution beyond the anti-adiabatic limit is through a tailored variational approach.  This typically requires some educated guess to the wavefunction.  In this context, our results act as a guess to the variational calculation.  On the other hand, a renormalization group approach can perhaps capture the arguments we presented above providing an insight into a solution in the adiabatic limit.  We leave all such efforts to future work.
}}

\section{EXPERIMENTAL CONSIDERATIONS}

\subsection{Ultracold quantum simulators}
The Peierls/SSH model can be realized in a variety of tight-binding models by implementing out-of-phase breathing-mode lattice distortions. 
Most promising are the experimental scenarios with polar molecules or cold atoms dressed by Rydberg states in optical lattices. 
In this case, the hard-core bare particles are the rotational excitations (for molecules) or electronic excitations (for cold atoms). 
If the molecules/atoms are trapped in a Mott insulator phase \cite{magneticfeshbach}, the excitations can be transferred between different sites of an optical lattice 
through couplings mediated by the dipole-dipole interactions. The transfer of such excitations between sites of an optical lattices have ben observed for molecules in Ref. \cite{PolarMolecules_Exp} and for Rydberg atoms in Ref. \cite{rydbergs5}. 

The bosonic field is provided by the translational motion of the trapped species in the lattice potential. Under such conditions, the translational excitations are nearly dispersionless Einstein optical lattice phonons coupled to the internal excitations of molecules/atoms due to the radial dependence of the dipolar interactions. This was illustrated for polar molecules in Ref. \cite{polar-molecules1,polar-molecules2} and for cold atoms dressed with Rydberg excitations in Ref. \cite{rydbergs3}.  In both of these cases, the hopping term couples to lattice distortions giving rise to a Peierls/SSH coupling $g= -3(t/a)\sqrt{\hbar/2m\Omega}$, where $m$ is the mass of the molecule/atom and $a$ is the lattice constant. The frequency of lattice phonons in a one-dimensional array is $\Omega = (2/\hbar) \sqrt{V_0E_R}$, where $V_0$ is the lattice depth and $E_R = \hbar^2 \pi^2 / 2ma^2$ is the recoil energy. The dimensionless Peierls/SSH coupling is then $\lambda = 18 E_R  t/(\hbar \pi \Omega)^2$. 
Thus the values of $\lambda$, which correspond to different points on the solid line in Fig. \ref{fig1}, can be obtained by varying $\Omega$.

In order to achieve the values of $\lambda > \lambda_C$ of the anti-adiabatic limit, one can either increase $t$, which can be done for cold atoms by dressing with Rydberg states, or decrease $\Omega$, which can be done for either atoms or molecules by simply decreasing the intensity of the optical lattice laser intensity.  In the latter case, care must be taken to prevent the Mott insulator phase from melting. The Mott insulator phase can be stabilized by inducing an on-site repulsion between molecules or atoms. This can be done either by tuning the scattering length of molecules or atoms by magnetic Feshbach resonances \cite{Feshbach_RMP} or by orienting molecules with weak DC fields to induce strongly repulsive dipolar interactions between molecules in the same internal state. It must be noted that the presence of a DC field may induce additional phonon-mediated interactions \cite{polar-molecules1,polar-molecules2}. Therefore, the field must be weak enough to ensure that the phonon-induced interactions discussed here remain dominant. 

The bipolaron appears as a sharp peak above the two-polaron continuum. Therefore, any measurement of the spectral function acts as a probe for the bipolaron.  An angle-resolved photoemission spectroscopy (ARPES)-like measurement scheme is well suited for this purpose.  To this end, the stimulated Raman spectroscopy scheme proposed in Ref. \cite{polar-molecules2} can be adopted to measure the two-particle spectral function $A(K,\omega) = -Im[G(k,1,\omega)]/\pi$.  

\subsection{Quantum materials}
Here we point out the fundamental importance of the Peierls/SSH model for quantum materials.  The Peierls/SSH model takes into consideration the linear term in the expansion of the hopping: $t({\bf R}_i - {\bf R}_j) \approx t + \alpha ({\bf R}_i - {\bf R}_j)$ around the lattice equilibrium position. Here, $t$ represents the hopping amplitude at equilibrium positions, ${\bf R}_i$ labels the position operator of lattice site $i$, and $\alpha$ depends on the lattice parameters $\Omega$ and $M$, the oscillator mass.  Second quantizing the ${\bf R}$ operators (and defining $g$ in terms of the lattice parameters) yields the SSH model with the coupling in Eq. (\ref{phonon-coupling}).

This interaction describes the coupling of the particle's motion to breathing-mode lattice distortions and represents linear corrections to the Born-Oppenheimer approximation in solids.  Even at low temperatures, zero-point motion may lead to drastic changes in the quasiparticle behavior if the coupling $g$ is comparable to the hopping amplitude $t$.  It is therefore important to distinguish effects of the Peierls/ SSH coupling in quantum materials and identify their experimental signatures. This model for hard-core particles can be applied to excitons in semiconductors and two-dimensional materials.  Alternatively, it can be applied to triplet pairing of fermionic excitations.  In all these materials, the repulsive bipolaron spectral peak, if present, must broaden due to coupling to low-energy phonons and will manifests itself as a resonance in a continuum.

To study the stability of the bipolaron in these systems, it would be interesting to develop an open quantum system description with couplings to dissipative channels given by the Peierls Hamiltonian.  We leave the study of these effects to future work, which will focus on variational approaches to the problem in a broader region of parameter space.

\section{CONCLUSION}
The repulsive Peierls/SSH bipolaron represents a novel quasiparticle that can be potentially realized with ultracold quantum simulators. It does not have a direct analog in quantum materials as the presence of acoustic phonons should embed such states into a strongly dissipating environment. The  lifetime of this quasiparticle in quantum materials would be finite.  However, it can leave a signature as a resonance in a continuum.  

The repulsive bipolaron discussed here is different from typical repulsively bound pairs in $t-V$ models.  The repulsive bipolaron is bound by phonon-mediated density-density and pair-hopping interactions. Experimental observations of field/bath-mediated interactions is fundamental to understanding complex phenomena in coupled field theories.  Additionally, the bipolaron possesses both pair-hopping and NNN hopping which typical repulsively bound pairs do not exhibit.  This means that quantum interference effects in quantum walks of the repulsive bipolaron ought to be different from those of $t-V$ or extended Hubbard repulsively bound states.  The repulsive bipolaron has a significantly smaller effective mass than the Hubbard repulsively bound pair owing to the pair-hopping kinetic terms. 

Our proposal represents an interesting mechanism for realizing repulsive interactions between pseudospins or excitons. Generally, excitons interact via dynamical interactions that are attractive owing to the specific tensorial form of the resonant dipole-dipole coupling \cite{agranovich}. Under conditions discussed here, excitons would interact via a phonon-mediated repulsive interaction giving rise to a Frenkel biexciton \cite{Frenkel1,Frenkel2,FrenkelBiexciton} dressed with phonons. Note that Frenkel biexcitons have never been observed in quantum materials because the hopping of the excitations is determined by the dipolar interactions between molecules, whereas the dynamical interactions between the excitations are determined by higher-order ({\it e.g.} quadrupolar) interactions due to the symmetry of the molecular states. The mechanism introduced here could be the leading mechanism for pairing of excitons in materials such as solid molecular hydrogen \cite{SolidH}, possessing high energy phonon modes. 

This emergent phonon-mediated repulsion can be used as a tunable parameter in quantum simulators. Applications can range from stabilizing pre-associated repulsively bound pairs in models with bare repulsive interactions to realization of spin models with pair-hopping and repulsion. For instance, the extension of our effective model to frustrated lattices such as triangular lattices would enable the study of frustration in spin liquids \cite{SpinLiquids} and supersolids \cite{Supersolids}. Studies of a frustrated model closely related to ours reveals a supersolid phase in one dimension \cite{Supersolid_1D} and on the triangular lattice \cite{Supersolid_triangle}. The persistence of phonon-mediated interactions at finite concentrations and in higher dimensions is vital for this research direction.  We note that the phonon-mediated NN repulsion is a result of statistics blocking hopping to NN sites. Thus, it is likely that the effect will persist in ensembles of greater number of particles, as the more confinement the particles experience, the more likely they are forced to be NN and interact via the repulsive mechanism discussed here. Phonon-mediated pair-hopping is also likely to survive in the anti-adiabatic limit. This is because a single phonon virtual excitation can allow for hopping of at most a pair of NN particles excluding ``cluster-hopping'' of ensembles of three, four and higher number of neighbouring particles.  A systematic approach must be developed to analyze all such terms and potential corrections at finite concentrations.

\begin{acknowledgments}
J.S. is grateful to Kirk Madison, Richard Schmidt and Francesco Scazza for enlightening discussions. This work was supported by the Natural Sciences and Engineering Research Council of Canada (NSERC) (J.S., R.V.K. and M.B.), the Stewart Blusson Quantum Matter Institute (SBQMI) (J.S. and M.B.) and the National Science Foundation (NSF) through a grant for the Institute for Theoretical Atomic, Molecular, and Optical Physics (ITAMP) at Harvard University and the Smithsonian Astrophysical Observatory in the form of an ITAMP student visitor fellowship (J.S.). \end{acknowledgments}

\appendix

\section{DERIVATION OF THE EFFECTIVE HAMILTONIAN IN THE ANTI-ADIABATIC LIMIT}
\label{PT}

Our method is inspired by the technique developed in Ref. \cite{Projection_technique}.  This technique provides an effective Hamiltonian description for the dressed particles.  Formally, the polaron operators in the full Hilbert space are mimicked by the action of the bare particle operators in the low-energy subspace, if the Hamiltonian is appropriately modified to include the relevant particle-phonon corrections. Below we outline the procedure.

\subsection{Single-bare-particle sector}

Let $\hat{P}$ be the projector onto the zero-phonon Hilbert subspace, which is spanned by the states $c_i^\dagger|0\rangle$, $\forall i$. The effective Hamiltonian in this subspace is, to second order:
\begin{equation*}
\EqLabel{s1}
\hat{h}_1 = \hat{T} + \hat{P} \hat{V} {1\over {E_0 - {\cal H}_0}} \hat{V}\hat{P},
\end{equation*}
where $\hat{T} =-t \sum_{i}^{} (c_i^\dagger c_{i+1} + h.c.)$ is the bare kinetic energy, $\hat{V}$ is the bare particle-phonon coupling from Eq. (\ref{phonon-coupling}), and ${\cal H}_0 = {\cal H}_{\rm ph}$. The projection is straightforward to carry out and leads to:
\begin{equation}
\EqLabel{s2}
\hat{h}_1 = \hat{T} - {4g^2\over \Omega} \sum_{i}^{} \hat{n}_i + \hat{T}_2,
\end{equation}
where $\hat{T}_2 = + t_2 \sum_{i}^{} (c_i^\dagger c_{i+2} + h.c.)$ is a phonon-mediated NNN hopping with $t_2 = g^2/\Omega= \lambda t/2$ (note the unusual sign). For convenience we define $\epsilon_0 ={ 4g^2\over \Omega}$.

The polaron dispersion is, therefore,  $$E_P(k) = - \epsilon_0 - 2t \cos(k) + 2 t_2 \cos(2k).$$ It is straightforward to verify that if $t > 4t_2$, {\em i.e.} if $\lambda < {1\over 2}$, the polaron ground state (GS) momentum is 0. For $\lambda > {1\over 2}$, the polaron GS momentum is $k_P= \arccos {t\over 4 t_2}$, going asymptotically to ${\pi \over 2}$ as $\lambda\rightarrow \infty$. 

\subsection{Two-bare-particle sector}

Repeating the projection onto the two-bare particle -- zero-phonon subspace spanned by the states $c^\dagger_i c^\dagger_{i+n}|0\rangle$, $\forall n \ge 1, i$, we find:
\begin{equation}
\EqLabel{s3}
\hat{h}_2 = - \epsilon_0\sum_{i}^{} \hat{n}_i + \hat{T} + \hat{T}_2 + V \sum_{i}^{} \hat{n}_i \hat{n}_{i+1},
\end{equation}
where $V = \epsilon_0$.  Note that this term simply adds to any bare NN interaction. The hard-core condition excludes processes where a particle hops past the other via $\hat{T}_2$.  Thus, $\hat{T}_2$ represents the pair-hopping of NN pairs as a whole.  See Ref.\cite{Repulsive} for more details.

\section{BOGOLIUBOV-BORN-GREEN-KIRKWOOD-YVON (BBGKY) EQUATION-OF-MOTION (EOM)}
\label{EOMBBGKY}

The repulsive bipolaron bound state dispersion can be calculated using the BBGKY EOM approach \cite{CiniSawatzky, MBmp}, which we outline below.

We define 
\begin{eqnarray}
|K,n\rangle = \sum_{i}^{} {e^{iK(R_i+{n\over 2})}\over \sqrt{N}} c^\dagger_i c^\dagger_{i+n}|0\rangle, \nonumber
\end{eqnarray}
$\forall n \ge 1$, and the propagators $G(K,n,\omega)=\bra{K,1}\hat{G(\omega)}\ket{K,n}$, where $\hat{G}(\omega) = ( \omega + i \eta - \hat{h}_2)^{-1}$ is the resolvent of interest. We use  the short-hand notation $g(n)$ for $G(K,n,\omega)$.  The bound state energy (once a bound state appears) is at the highest discrete pole of these propagators above the two-polaron continuum.  Using the identity $\hat{G}(\omega) ( \omega + i \eta - \hat{h}_2)^{-1}=1$, we generate the EOM:
\begin{eqnarray}
&&(\omega + i \eta +2\epsilon_0-\beta_K -V) g(1)  = 1 - \alpha_K g(2) + \beta_K g(3) \nonumber \\
&&(\omega + i \eta +2\epsilon_0) g(2)   = -\alpha_K [g(1)+g(3)]+ \beta_K g(4) 
\end{eqnarray}
and for any $n\ge 3$,
\begin{eqnarray}
   (\omega + i \eta +2\epsilon_0) g(n)   = &&-\alpha_K [g(n-1)+g(n+1)] \nonumber\\ &&+ \beta_K [g(n-2)+g(n+2)].
\end{eqnarray}
   Here,  $\alpha_K = 2t \cos ({K\over 2})$, $\beta_K = 2t_2 \cos(K)$.

 The physically acceptable analytical solution for recurrence
 relations of this type is available in \cite{Mirko_Math}, however it is rather
 complicated and its poles cannot be extracted analytically. A general
 solution can be found numerically.  Results in Fig. \ref{fig1} represent solutions obtained by solving this system of equations numerically. 
 
We note that for $K=\pi$, $\alpha_K$ vanishes limiting the EOM to odd $n$.  This explains why even $n$ states are forbidden as shown in the inset of Fig. \ref{fig2}b).

\bibliography{RepulsiveBipolaron}

\end{document}